\documentclass[aps,prl,reprint,nofootinbib,superscriptaddress]{revtex4-2}

\usepackage[utf8]{inputenc}
\usepackage[T1]{fontenc}
\usepackage{amsmath,amssymb,amsfonts,bm}
\usepackage{mathtools}
\usepackage{hyperref}
\usepackage{xcolor}

\hypersetup{
colorlinks=true,
linkcolor=blue,
citecolor=blue,
urlcolor=blue
}

\newcommand{\ii}{\mathrm{i}}
\newcommand{\ee}{\mathrm{e}}

\begin{document}

\title{State-resolved quantum transport of vortex electrons in accelerators}

\author{S.~S.~Baturin}
\email{s.s.baturin@gmail.com}
\affiliation{School of Physics and Engineering, ITMO University, St. Petersburg 197101, Russia}

\date{\today}

\begin{abstract}
Vortex electrons carry a quantized orbital angular momentum (OAM) degree
of freedom, but whether this internal structure can survive repeated
transport through an accelerator lattice remains unclear. Here we
formulate a density-matrix theory for periodic round lattices and show
that the symmetry-protected quantity is a Lewis-Floquet OAM invariant,
rather than the instantaneous kinetic OAM. The classical transfer map
lifts to unitary state evolution, while stochastic field errors generate
a Lindblad channel. This framework exposes a sharp separation between
visibility and state survival. Dipole jitter displaces the wavepacket,
rapidly smearing a vortex signature measured about a fixed origin
without altering its recentered internal OAM distribution. Quadrupole
fluctuations instead drive genuine $\Delta\ell=\pm2$ leakage. For a
matched $n=0$, $|\ell|=1$ mode, white-noise estimates based on
representative IOTA and PETRA III parameters give fixed-frame smearing
scales of $4.9\times10^2$ and $3.3$ turns, but intrinsic-leakage scales
of $2.1\times10^5$ and $2.2\times10^6$ turns, respectively. Thus loss
of an unrecentered vortex image need not signal destruction of the
vortex state: its internal OAM structure can persist hundreds to
hundreds of thousands of times longer. Centroid tracking and quadrupole
stability are therefore distinct experimental requirements for
observability and state survival, respectively.
\end{abstract}

\maketitle

Vortex electrons are structured matter waves whose phase winds around
the propagation axis and whose transverse state carries quantized
intrinsic orbital angular momentum (OAM). This additional degree of
freedom has stimulated developments across electron optics, relativistic
wave mechanics, and high-energy applications
\cite{Bliokh2007,Bliokh2011,BliokhReview}. Transporting such states
through a linear accelerator or storing them for many revolutions would
extend vortex-electron physics to relativistic energies and repeated
interactions. It also poses a basic question: can an accelerator preserve
the full orbital quantum state, rather than only its average angular
momentum?

Relativistic descriptions have established how the mean angular
momentum of vortex particles evolves in external fields and accelerator
systems \cite{Silenko2018,Karlovets2019,Karlovets2026}. The mean alone,
however, does not determine whether a vortex state is preserved.
Populations can leak between orbital modes, coherences can decay, and a
pure state can become mixed while retaining the same expectation value.
Conversely, a displaced but otherwise intact wavepacket can appear to
lose its vortex signature when measured about a fixed laboratory origin.
A state-resolved treatment must therefore connect the transfer maps of
accelerator optics to the evolution of the complete density matrix and
distinguish loss of observability from loss of the internal OAM state.

This distinction is nontrivial because the instantaneous kinetic OAM is
not generally conserved in a curved, alternating-gradient lattice as its
local evolution couples to the other quadratic degrees of freedom. Periodicity nevertheless
allows a different form of protection. The circular-mode, or round-beam,
lattice introduced by Burov, Nagaitsev, and Derbenev
\cite{Burov2002} possesses the transverse rotational symmetry needed to
define a stroboscopic OAM invariant. Lewis-Riesenfeld and Floquet theory
then provide its periodic continuation \cite{Lewis,Lewis1,Lewis2}, while
the metaplectic lift of the classical symplectic map promotes the usual
coordinate and covariance transport to unitary evolution of the full
quantum state \cite{Folland1989,deGosson2011}.

Real machines also contain orbit and injection jitter, magnet-gradient
fluctuations, rolls, radio-frequency noise, and deterministic mismatch.
These perturbations need not have the same physical consequence. A
known offset or optical mismatch drives coherent and, in principle,
reversible motion relative to the matched basis. Stochastic errors
instead produce ensemble decoherence and are naturally described as an
open quantum channel \cite{Lindblad1976,GKS1976,Breuer2007,Holevo2019}.
The relevant classification is set by the angular harmonic of each
perturbation: dipole terms translate the centroid, whereas quadrupole
and higher multipoles can resolve and redistribute the internal orbital
structure.

Here we develop a density-matrix theory of vortex-electron transport in
a periodic round lattice. We identify the Lewis-Floquet OAM invariant
that defines the natural state basis, construct the unitary ideal
transport, and derive the Lindblad channel generated by stochastic
machine errors. The resulting selection rules show that dipole jitter
creates potentially removable fixed-frame smearing, while quadrupole noise is the
leading intrinsic channel and transfers population by
$\Delta\ell=\pm2$. We derive initial leakage rates for individual and
correlated magnet errors and evaluate them for a matched $n=0$,
$|\ell|=1$ channel using representative IOTA and PETRA III machine
scales. We find a pronounced separation. Namely, the fixed-frame signatures
have smearing scales of $4.9\times10^2$ and $3.3$ turns, respectively,
whereas the corresponding intrinsic-leakage scales are
$2.1\times10^5$ and $2.2\times10^6$ turns. The disappearance of an
unrecentered vortex image can therefore precede destruction of the
internal OAM state by factors ranging from hundreds to hundreds of
thousands. Thus we conclude that centroid tracking
governs observability, whereas quadrupole perturbations affects genuine
state survival. Although we focus on a storage ring, the finite-map
construction also applies to a linear accelerator.

\textit{Relativistic ring Hamiltonian and quadratic operator dynamics.}
Let $\mathbf R_0(s)$ be the reference closed orbit and
$\{\mathbf e_x,\mathbf e_y,\mathbf e_s\}$ the associated Frenet frame.
Before passing to the paraxial ring description, we use the
spin-independent relativistic single-particle Hamiltonian \cite{Lee2018}
\begin{align}
\hat H_{\rm rel}
&=
q\Phi(\hat{\mathbf r},t)
+
\sqrt{
m^2c^4
+
c^2
\left[
\hat{\mathbf P}
-
q\mathbf A(\hat{\mathbf r},t)
\right]^2
}.
\label{eq:main_relativistic_hamiltonian}
\end{align}
We note that spin effects can be included as perturbations \cite{Gatalina2026,Epov2026}.
For the ring dynamics we normalize the canonical momenta by the reference
momentum $p_0$ and use
\begin{align}
\hat p_j
&=
\frac{\hat P_j}{p_0},
&
\hbar_{\rm eff}
&=
\frac{\hbar}{p_0}
\equiv
\bar\lambda_e.
\label{eq:main_effective_planck_constant}
\end{align}
The $s$-evolution equation is therefore
$\ii\hbar_{\rm eff}\partial_s\lvert\psi\rangle=
\hat H_F\lvert\psi\rangle$. Expanding about the reference orbit gives,
to second order in the centered normalized canonical variables, the
standard quadratic ring Hamiltonian
\cite{Lee2018,Dragt2019},
\begin{align}
\hat H_F^{(2)}(s)
&=
\frac{1}{2}
\hat Z^{T}h_F(s)\hat Z,
&
\hat Z
&=
\left(
\hat x,
\hat y,
\hat\zeta,
\hat p_x,
\hat p_y,
\hat p_\zeta
\right)^{T},
\label{eq:main_ring_hamiltonian}
\end{align}
where $h_F(s)=h_F^{T}(s)$ is a real symmetric matrix containing the
local curvature, guide field, quadrupole focusing, dispersion, RF
couplings, and linearized synchro-betatron terms.

Introduce the canonical symplectic matrix
\begin{align}
\mathsf J_6
&=
\begin{pmatrix}
0 & \mathsf I_3\\
-\mathsf I_3 & 0
\end{pmatrix},
&
\mathsf A_F(s)
&=
\mathsf J_6 h_F(s).
\label{eq:main_hamiltonian_matrix}
\end{align}
The canonical commutators are
$[\hat Z_a,\hat Z_b]=\ii\hbar_{\rm eff}(\mathsf J_6)_{ab}$.
The Heisenberg equations for the canonical operators then close linearly,
\begin{align}
\frac{d\hat Z}{ds}
&=
\mathsf A_F(s)\hat Z.
\label{eq:main_linear_heisenberg}
\end{align}
The linear evolution of $\hat Z$ induces a closed evolution on the full
algebra of quadratic operators. For any real symmetric matrix
$g(s)=g^{T}(s)$, define
\begin{align}
\hat{\mathcal Q}[g(s)]
&=
\frac{1}{2}
\hat Z^{T}g(s)\hat Z.
\label{eq:main_general_quadratic_operator}
\end{align}
Its total Heisenberg derivative is
\begin{align}
\frac{d}{ds}\hat{\mathcal Q}[g(s)]
&=
\frac{1}{2}
\hat Z^{T}
\left[
g'(s)
+
\mathsf A_F^{T}(s)g(s)
+
g(s)\mathsf A_F(s)
\right]
\hat Z.
\label{eq:main_quadratic_operator_evolution}
\end{align}
Thus the covariance matrix familiar from accelerator physics is only the
expectation-value projection of a closed operator algebra,
\begin{align}
\Sigma_{ab}(s)
&=
\frac{1}{2}
\left\langle
\hat Z_a\hat Z_b+\hat Z_b\hat Z_a
\right\rangle
-
\left\langle\hat Z_a\right\rangle
\left\langle\hat Z_b\right\rangle.
\label{eq:main_covariance}
\end{align}
The normalized representative of the physically measurable longitudinal
kinetic OAM is
\begin{align}
\hat{\mathcal L}_s(s)
&=
\hat x\hat\pi_y-\hat y\hat\pi_x,
&
\hat{\boldsymbol\pi}
&=
\hat{\mathbf p}-\frac{q}{p_0}\mathbf A,
\label{eq:main_ls}
\end{align}
so that the physical operator is
$\hat L_s^{\rm kin}=p_0\hat{\mathcal L}_s$. In particular, an eigenvalue
$\ell\hbar_{\rm eff}$ of the normalized OAM corresponds to the physical
value $\ell\hbar$.
To quadratic order about the reference orbit, it can be represented as
\begin{align}
\hat{\mathcal L}_s(s)
&=
\hat{\mathcal Q}
\left[
g_{\mathcal L}(s)
\right].
\label{eq:main_ls_quadratic_form}
\end{align}
The explicit matrix representation is given in the Supplemental
Material. 
Its $s$ dependence contains the
local vector potential, the rotation of the Frenet frame, and the
relation between canonical and kinetic momenta. Its exact quadratic-order
evolution is therefore
\begin{align}
\frac{d\hat{\mathcal L}_s}{ds}
&=
\frac{1}{2}
\hat Z^{T}
\left[
g_{\mathcal L}'(s)
+
\mathsf A_F^{T}(s)g_{\mathcal L}(s)
+
g_{\mathcal L}(s)\mathsf A_F(s)
\right]
\hat Z.
\label{eq:main_oam_nonclosed}
\end{align}
In a generic ring, the matrix in brackets does not vanish and is not
proportional to $g_{\mathcal L}(s)$ itself. Consequently, the evolution
of the local kinetic OAM is not closed on $\hat{\mathcal L}_s$. It is
coupled to the remaining quadratic operators, including the isotropic
envelope sector and the traceless quadrupolar shear sector.

This observation also separates local transport from one-turn
conservation. Let
\begin{align}
\mathsf M_C
&=
\mathsf M(s_0+C,s_0)
\label{eq:main_one_turn_map}
\end{align}
be the one-turn map at a fixed observation point $s_0$. A fixed
quadratic operator
\begin{align}
\hat{\mathcal I}
&=
\hat{\mathcal Q}[g_{\mathcal I}]
\end{align}
is conserved stroboscopically if and only if
\begin{align}
\mathsf M_C^{T}
g_{\mathcal I}
\mathsf M_C
&=
g_{\mathcal I}.
\label{eq:main_stroboscopic_matrix_condition}
\end{align}
At the operator level, this condition is equivalent to
\begin{align}
\hat U_C^{\dagger}
\hat{\mathcal I}
\hat U_C
&=
\hat{\mathcal I},
&
\hat U_C
&=
\hat U(s_0+C,s_0).
\label{eq:main_stroboscopic_invariant}
\end{align}
Equation~\eqref{eq:main_stroboscopic_invariant} is a stroboscopic
conservation law for a periodic ring.

For a round transverse one-turn map, one may introduce normalized
canonical variables
\begin{align}
\hat Z_n
&=
\left(
\hat X,
\hat Y,
\hat P_X,
\hat P_Y
\right)^{T}
\end{align}
such that the normalized map commutes with simultaneous rotations of
the transverse coordinates and momenta,
\begin{align}
\left[
\mathsf M_{n,C},
\mathsf R(\theta)
\right]
&=
0,
&
\mathsf R(\theta)
&=
\begin{pmatrix}
R_2(\theta) & 0\\
0 & R_2(\theta)
\end{pmatrix},
\label{eq:main_round_map_condition}
\end{align}
where
\begin{align}
R_2(\theta)
&=
\begin{pmatrix}
\cos\theta & -\sin\theta\\
\sin\theta & \cos\theta
\end{pmatrix}.
\end{align}
This is the operator form of the round-beam, or circular-mode, lattice
condition introduced by Burov, Nagaitsev, and Derbenev
\cite{Burov2002}. The generator of these normalized rotations is
\begin{align}
\hat L_n
&=
\hat X\hat P_Y-\hat Y\hat P_X.
\label{eq:main_normalized_oam}
\end{align}
The round-map condition implies
\begin{align}
\hat U_C^{\dagger}
\hat L_n
\hat U_C
&=
\hat L_n.
\label{eq:main_round_oam_invariant}
\end{align}
Thus $\hat L_n$ is an exact stroboscopic quadratic invariant of the
ideal round one-turn map. This operator statement holds independently
of whether the transported state is matched or mismatched.

The one-turn invariant admits a continuous formulation in terms of the
Lewis-Riesenfeld theory \cite{Lewis,Lewis1,Lewis2}. For the periodic ring Hamiltonian,
$\hat H_F^{(2)}(s+C)=\hat H_F^{(2)}(s)$, we introduce a Hermitian
quadratic operator $\hat{\mathcal I}_L(s)$ satisfying
\begin{align}
\frac{\partial\hat{\mathcal I}_L(s)}{\partial s}
+
\frac{\ii}{\hbar_{\rm eff}}
\left[
\hat H_F^{(2)}(s),
\hat{\mathcal I}_L(s)
\right]
&=
0,
\nonumber\\
\hat{\mathcal I}_L(s+C)
&=
\hat{\mathcal I}_L(s).
\label{eq:main_periodic_lewis_invariant}
\end{align}
The first relation is the Lewis-Riesenfeld invariant equation, while
the second selects the periodic, or Floquet, representative appropriate
to a storage ring. At a fixed observation point $s_0$, these conditions
imply
\begin{align}
\hat U_C^{\dagger}
\hat{\mathcal I}_L(s_0)
\hat U_C
&=
\hat{\mathcal I}_L(s_0),
&
\hat U_C
&=
\hat U(s_0+C,s_0).
\label{eq:main_lewis_stroboscopic_invariant}
\end{align}
Thus $\hat{\mathcal I}_L(s)$ is transported continuously as a
Lewis-Floquet invariant, whereas its representative at a fixed
azimuth is conserved stroboscopically by the one-turn propagator.

The stable round one-turn map admits two commuting quadratic invariants:
the total oscillator-action invariant $\hat{\mathcal I}_0(s)$ and the
Lewis-Floquet OAM invariant $\hat{\mathcal I}_L(s)$. We define their
common eigenbasis by
\begin{align}
\hat{\mathcal I}_0(s)\lvert\phi_{n\ell}(s)\rangle
&=\hbar_{\rm eff}(2n+|\ell|+1)
\lvert\phi_{n\ell}(s)\rangle, \nonumber\\
\hat{\mathcal I}_L(s)\lvert\phi_{n\ell}(s)\rangle
&=\hbar_{\rm eff}\ell\lvert\phi_{n\ell}(s)\rangle,
\end{align}
where $n\in\mathbb N_0$ and $\ell\in\mathbb Z$. The corresponding
circular actions are the derived combinations
$\hat{\mathcal I}_{\pm}
=(\hat{\mathcal I}_0\pm\hat{\mathcal I}_L)/2$.

The eigenstates $\lvert\phi_{n\ell}(s)\rangle$ define a stroboscopic Lewis-Floquet basis at the
chosen azimuth of the ring and reproduce themselves after one circumference,
while the exact solutions acquire the associated Lewis-Riesenfeld, or
equivalently Floquet, phases. 


\textit{Metaplectic lift of the ideal ring dynamics.}
We note that the Heisenberg equations generated by
Eq.~\eqref{eq:main_ring_hamiltonian} are linear, so the same classical
transport that determines the periodic Lewis-Floquet invariants is represented
by a symplectic matrix $\mathsf M(s,s_0)$. Every such map admits a
metaplectic lift $\hat U(s,s_0)$, unique up to the double-cover
sign, such that \cite{Folland1989,deGosson2011}
\begin{align}
\hat U^\dagger(s,s_0)\hat Z\hat U(s,s_0)
&=
\mathsf M(s,s_0)\hat Z,
\nonumber\\
\rho(s)
&=
\hat U(s,s_0)\rho(s_0)\hat U^\dagger(s,s_0).
\label{eq:main_metaplectic}
\end{align}
The corresponding transport of each quadratic Lewis-Floquet invariant is
\begin{align}
\hat{\mathcal I}_a(s)
&=
\hat U(s,s_0)\hat{\mathcal I}_a(s_0)
\hat U^\dagger(s,s_0),
&
a&=0,L.
\label{eq:main_metaplectic_lewis_transport}
\end{align}
Equation~\eqref{eq:main_metaplectic_lewis_transport} is equivalent to
the Lewis-Riesenfeld equation
\eqref{eq:main_periodic_lewis_invariant}.  At one turn, periodicity of
the invariant gives the exact relation
\begin{align}
\hat U_C^\dagger\hat{\mathcal I}_a(s_0)\hat U_C
&=
\hat{\mathcal I}_a(s_0),
&
\hat U_C
&=
\hat U(s_0+C,s_0).
\label{eq:main_metaplectic_lewis_one_turn}
\end{align}
Thus the covariance law is only the second-moment shadow of the same
unitary transport that carries the complete Lewis-Floquet basis.  


In particular, let $\Pi_\ell$ denote the spectral projector of
$\hat{\mathcal I}_L(s_0)$ labeled by
$\ell\in\mathbb Z$.  The ideal one-turn evolution preserves
\begin{align}
P_\ell(N)
&=
\operatorname{Tr}\!\left[
\Pi_\ell\hat U_C^N\rho(s_0)\hat U_C^{\dagger N}
\right]
=
P_\ell(0).
\label{eq:main_ideal_oam_spectrum}
\end{align}
This is the state-level meaning of symmetry protection.  The protected
operator is the periodic Lewis-Floquet invariant
$\hat{\mathcal I}_L(s_0)$, not the instantaneous kinetic OAM
$\hat{\mathcal L}_s(s)$, which need not commute with the local
Hamiltonian and may exchange continuously with the remaining
quadratic sector.

A matched basis mode is a single Lewis-Floquet eigenstate
$\lvert\phi_{n\ell}(s)\rangle$, and a density matrix diagonal in this
basis is stroboscopically stationary. A
deterministic mismatch is instead a coherent affine-symplectic
transformation relative to that basis. Its centroid part is a Weyl
displacement $\hat W(Z_c)$, while breathing and shear are described by
$\hat U_{\rm mis}$. At the state
level one may write
\begin{align}
\rho_{\rm mis}(s_0)
&=
\hat W(Z_c)\hat U_{\rm mis}
\rho_{\rm mat}(s_0)
\hat U_{\rm mis}^\dagger\hat W^\dagger(Z_c),
\nonumber\\
\hat U_{\rm mis}
&=
\exp\!\left[
-\frac{\ii}{2\hbar_{\rm eff}}\hat Z^T g_{\rm mis}\hat Z
\right],
\label{eq:main_mismatch_operator}
\end{align}
where the corresponding coherent Hamiltonian contains both linear and
quadratic terms,
$\hat H_{\rm mis}=\mathbf f_{\rm mis}^{T}\hat Z+
\tfrac12\hat Z^{T}h_{\rm mis}\hat Z$. Thus a known mismatch is
reversible and may be included in the target state. A scalar radial
squeeze commutes with
$\hat{\mathcal I}_L$, whereas traceless shear does not and therefore
produces coherent motion between circular and astigmatic sectors that leads to mode conversion
\cite{Padgett1999,Calvo2005,Cisowski2022,Filina2023,Filina2026}.

The same lift defines the natural interaction picture for the open
dynamics,
\begin{align}
\widetilde\rho(s)
&=
\hat U^\dagger(s,s_0)\rho(s)\hat U(s,s_0).
\label{eq:main_ideal_interaction_picture}
\end{align}
We reiterate that in this frame the ideal Lewis-Floquet transport is removed and the
ideal density matrix is stationary. Deterministic mismatch remains a
coherent affine-quadratic perturbation, while stochastic imperfections
act through the correspondingly transformed generators. The Lindblad
construction below therefore measures diffusion relative to the
natural ring basis selected by the Lewis invariants.

\textit{Imperfections: coherent mismatch and stochastic noise.}
To describe a realistic machine, we separate deterministic mismatch from stochastic imperfections and write
\begin{align}
\hat H(s)
&=
\hat H_0(s)+\hat H_{\rm mis}(s)+\delta\hat H(s),
\nonumber\\
\delta\hat H(s)
&=
\sum_\alpha \xi_\alpha(s)\hat G_\alpha(s),
\label{eq:main_noise_hamiltonian}
\end{align}
where $\hat H_0$ is the ideal matched round-lattice Hamiltonian, $\hat H_{\rm mis}$ is the deterministic mismatch Hamiltonian, and $\delta\hat H$ is the stochastic part. The mismatch term describes coherent departures from the matched round mode, including centroid offsets, breathing mismatch, shear (astigmatic) mismatch, and higher deterministic multipolar distortions. The stochastic processes $\xi_\alpha$ describe orbit jitter, injection jitter, gradient errors, rolls, RF noise, and nonlinear field imperfections, while the Hermitian operators $\hat G_\alpha$ are the corresponding generators.

For zero-mean Markovian noise,
\begin{equation}
\mathbb E[\xi_\alpha(s)\xi_\beta(s')]
=
D_{\alpha\beta}(s)\delta(s-s'),
\end{equation}
where $D_{\alpha\beta}(s)$ is the positive semidefinite correlation matrix in imperfection-channel space. The ensemble-averaged density matrix then obeys
\begin{align}
\frac{d\bar\rho}{ds}
&=
-\frac{\ii}{\hbar_{\rm eff}}
[\hat H_0(s)+\hat H_{\rm mis}(s),\bar\rho]
\nonumber\\
&\quad
-\frac{1}{2\hbar_{\rm eff}^2}
\sum_{\alpha\beta}
D_{\alpha\beta}(s)
[\hat G_\alpha(s),[\hat G_\beta(s),\bar\rho]].
\label{eq:main_lindblad}
\end{align}
After diagonalizing $D_{\alpha\beta}$, Eq.~\eqref{eq:main_lindblad} takes the standard Lindblad form with Hermitian Lindblad operators. In this decomposition, the mismatch remains part of the coherent Hamiltonian and produces reversible mode mixing, whereas the Lindblad term describes the genuinely stochastic part of the machine and hence irreversible decoherence.

Deterministic mismatch and stochastic imperfections are classified by
the same angular harmonics. For either
$\hat O=\hat H_{\rm mis}$ or $\hat O=\hat G_\alpha$, write
\begin{align}
\hat O
&=
\sum_m \hat O_m,
\nonumber\\
[\hat{\mathcal I}_L(s),\hat O_m(s)]
&=
m\hbar_{\rm eff}\,\hat O_m(s).
\label{eq:main_perturbation_harmonics}
\end{align}
Equivalently, in the interaction picture these commutators are taken
with the fixed operator $\hat{\mathcal I}_L(s_0)$.  The sectors labeled
by $\ell$ below are therefore the eigenspaces of the natural ring
invariant.
Hence both coherent and stochastic perturbations obey the same OAM selection rule:
\begin{equation}
\langle \ell',\nu'\rvert\hat O_m\lvert\ell,\nu\rangle \neq 0
\qquad
\text{only if}
\qquad
\ell'-\ell=m.
\label{eq:main_selection_rule_generic}
\end{equation}
Deterministic $m\neq0$ components produce coherent
oscillatory coupling, whereas stochastic $m\neq0$ components produce
population transfer and dephasing after ensemble averaging. A dipole
perturbation generates $\Delta\ell=\pm1$ sidebands about a fixed
laboratory origin, but it is only a displacement and disappears in the
centroid frame. Quadrupole perturbations generate
$\Delta\ell=\pm2$ sidebands that remain after recentering and therefore
constitute the leading intrinsic OAM channel.

\textit{Laboratory smearing and intrinsic OAM diffusion.}
To quantify the survival of a target pure state
$\rho_\psi=\lvert\psi\rangle\langle\psi\rvert$, we use the interaction
picture in Eq.~\eqref{eq:main_ideal_interaction_picture} and define
\begin{align}
F(s)
&=
\langle\psi\rvert\widetilde{\bar\rho}(s)\lvert\psi\rangle.
\label{eq:main_fidelity}
\end{align}
If initially $\widetilde{\bar\rho}(s_0)=\rho_\psi$,
Eq.~\eqref{eq:main_lindblad} gives the exact initial stochastic loss rate
\begin{align}
\Gamma_F(s_0)
&=
-\left.\frac{dF}{ds}\right|_{s=s_0}
\nonumber\\
&=
\frac{1}{\hbar_{\rm eff}^2}
\sum_{\alpha\beta}
D_{\alpha\beta}(s_0)
\operatorname{Cov}^{\mathrm{sym}}_\psi
(\hat G_{\alpha I},\hat G_{\beta I}),
\label{eq:main_fidelity_rate}
\end{align}
where $\hat G_{\alpha I}(s)=\hat U^\dagger(s,s_0)
\hat G_\alpha(s)\hat U(s,s_0)$
and
$\operatorname{Cov}^{\mathrm{sym}}(A,B)=
\langle\{A,B\}\rangle/2-\langle A\rangle\langle B\rangle$.
Deterministic mismatch changes this covariance but does not itself
produce irreversible loss.

Machine errors are usually specified as integrated kicks. For
independent zero-mean kicks renewed once per turn, let
$C_{ab}=\mathbb E[\delta g_a\delta g_b]$ be their one-turn covariance
and $\hat G_a$ the corresponding interaction-picture generators. The
initial loss per turn is
\begin{align}
\Lambda
&=
\frac{1}{\hbar_{\rm eff}^2}
\sum_{ab}C_{ab}
\operatorname{Cov}^{\mathrm{sym}}_\psi(\hat G_a,\hat G_b).
\label{eq:main_one_turn_loss}
\end{align}
For $\Lambda\ll1$, the no-return, or absorbing-channel, estimate is
$F_N\simeq\exp(-N\Lambda)$. We use
$N_{\rm ch}=\Lambda^{-1}$ as the characteristic channel lifetime. 

A thin dipole error at location $i$ gives
\begin{align}
\delta\hat V_{d,i}
&=
\delta\theta_{x,i}\hat x_i
+
\delta\theta_{y,i}\hat y_i.
\label{eq:main_dipole_kick}
\end{align}
For a matched circular state with $n=0$ and circular beta function
$\beta_i$,
\begin{align}
\operatorname{Var}(\hat x_i)
=
\operatorname{Var}(\hat y_i)
&=
\frac{|\ell|+1}{2}\,\beta_i\hbar_{\rm eff}.
\label{eq:main_circular_second_moment}
\end{align}
Independent angular kicks therefore give the fixed-frame loss
\begin{align}
\Lambda_d
&=
\frac{|\ell|+1}{2\hbar_{\rm eff}}\,
\mathcal A_d,
&
\mathcal A_d
&=
\sum_i\beta_i
\left(
\sigma_{\theta x,i}^2+
\sigma_{\theta y,i}^2
\right).
\label{eq:main_dipole_loss_per_turn}
\end{align}
This is a laboratory-observability scale. Each realization is a Weyl
displacement of the same internal state. If the density matrix is
recentered with the measured centroid, the dipole contribution vanishes
and the intrinsic OAM distribution is unchanged.

The leading intrinsic channel is a thin normal or skew quadrupole kick,
\begin{align}
\delta\hat V_{q,i}
&=
\delta K_{c,i}\hat Q_{c,i}
+
\delta K_{s,i}\hat Q_{s,i},
\nonumber\\
\hat Q_{c,i}
&=
\frac12(\hat x_i^2-\hat y_i^2),
&
\hat Q_{s,i}
&=
\hat x_i\hat y_i,
\label{eq:main_quadrupole_kick}
\end{align}
where $\delta K_{\lambda,i}=\int ds\,\delta k_{\lambda,i}$ has units
of inverse length. With $\hat r_i^2=\hat x_i^2+\hat y_i^2$, rotational
symmetry and
$\langle \hat r_i^4\rangle=(|\ell|+1)(|\ell|+2)
(\beta_i\hbar_{\rm eff})^2$ give
\begin{align}
\operatorname{Var}(\hat Q_{c,i})
=
\operatorname{Var}(\hat Q_{s,i})
&=
\frac{(|\ell|+1)(|\ell|+2)}{8}
(\beta_i\hbar_{\rm eff})^2.
\label{eq:main_quadrupole_variance}
\end{align}
Thus, for independent normal and skew errors,
\begin{align}
\Lambda_q
&=
\frac{(|\ell|+1)(|\ell|+2)}{8}
\sum_i\beta_i^2
\left(
\sigma_{Kc,i}^2+
\sigma_{Ks,i}^2
\right).
\label{eq:main_quadrupole_loss_per_turn}
\end{align}
Unlike Eq.~\eqref{eq:main_dipole_loss_per_turn}, this leakage remains in
the centroid frame and populates the $\ell\pm2$ sectors.

For normal-quadrupole current jitter,
$\sigma_{Kc,i}=|k_{1,i}|L_i\sigma_{I,i}$ and
$\sigma_{Ks,i}=0$, so
\begin{align}
\Lambda_q
&=
\frac{(|\ell|+1)(|\ell|+2)}{8}
\sum_i
\left(k_{1,i}\beta_iL_i\sigma_{I,i}\right)^2.
\label{eq:main_quadrupole_machine_loss}
\end{align}

We note that if normal and skew components of equal variance are
both present, the result doubles. For magnets connected to common
supplies, the diagonal sum must be replaced by the covariance quadratic
form
\begin{align}
\Lambda_q
&=
\frac{(|\ell|+1)(|\ell|+2)}{8}
\sum_{ab}C^{(I)}_{ab}\,
\mathbf v_a\!\cdot\!\mathbf v_b,
\nonumber\\
\mathbf v_a
&=
\sum_{i\in a}k_{1,i}\beta_iL_i
\left(
\cos2\varphi_i,
\sin2\varphi_i
\right),
\label{eq:main_correlated_quadrupole_loss}
\end{align}
where $\varphi_i$ is the shear phase supplied by the normalized round
transport and $C^{(I)}$ is the covariance of the fractional supply
errors. Equation~\eqref{eq:main_correlated_quadrupole_loss} is the form
to use with a measured lattice and noise spectrum.

\textit{IOTA and PETRA III estimates.}
In the following example we combine documented machine scales with
the noise assumptions for a matched round channel. IOTA operates with $100$--$150\,\mathrm{MeV}$
electrons in a $39.97\,\mathrm{m}$ ring and contains about forty
quadrupoles \cite{Antipov2017IOTA,Valishev2021IOTA}. PETRA III operates
at $6\,\mathrm{GeV}$ with circumference $2304\,\mathrm{m}$; its
published quadrupole ranges are
$k_1=0.749$--$1.049\,\mathrm{m}^{-2}$ and
$L=0.44$--$1.04\,\mathrm{m}$ \cite{DESYPETRAIIIParameters}. For an
uncorrelated-magnet estimate we use
$(N_{\rm quad},k_1,\beta,L)=(40,5\,\mathrm{m}^{-2},4\,\mathrm{m},
0.2\,\mathrm{m})$ for IOTA and
$(375,0.8\,\mathrm{m}^{-2},10\,\mathrm{m},0.5\,\mathrm{m})$ for
PETRA III. This gives
\begin{align}
\sum_i(k_{1,i}\beta_iL_i)^2
\simeq
\begin{cases}
6.4\times10^2, & \text{IOTA},\\
6.0\times10^3, & \text{PETRA III}.
\end{cases}
\label{eq:main_benchmark_optics_factors}
\end{align}
We further assume white, turn-uncorrelated fractional gradient noise
$\sigma_I=10^{-4}$ and $10^{-5}$ for IOTA and PETRA III, respectively.
For the dipole estimates we define one effective single-plane kick by
$\mathcal A_d=\beta_d\theta_{\rm eff}^2$ and set
$\theta_{\rm eff}=1\,\mathrm{nrad}$, with
$\beta_d=4\,\mathrm{m}$ for IOTA and $10\,\mathrm{m}$ for PETRA III.
For the $|\ell|=1$ modes Eq.~\eqref{eq:main_circular_second_moment} then gives rms widths
$\sigma_x=89\,\mathrm{nm}$ and $18\,\mathrm{nm}$, respectively.
The corresponding pairs $(\bar\lambda_e,T_0)$ are
$(1.96\,\mathrm{fm},133.3\,\mathrm{ns})$ and
$(3.29\times10^{-2}\,\mathrm{fm},7.685\,\mu\mathrm{s})$.
In Table~\ref{tab:main_machine_estim}, we summarize the estimated inverse initial leakage scales, expressed in turns, for the laboratory-observable signature of the stored vortex state, $N_{\rm lab}^{(d)}=\Lambda_d^{-1}$, and for its intrinsic OAM sector, $N_{\rm int}^{(q)}=\Lambda_q^{-1}$.

\begin{table}[t]
\caption{Laboratory smearing and intrinsic leakage scales for a matched
$n=0$, $|\ell|=1$ state. Times $N_{\rm ch}T_0$ are in parentheses.}
\label{tab:main_machine_estim}
\begin{ruledtabular}
\begin{tabular}{lcc}
Ring
& $N_{\rm lab}^{(d)}$
& $N_{\rm int}^{(q)}$\\
\hline
IOTA
& $4.9\times10^2\;(65\,\mu\mathrm{s})$
& $2.1\times10^5\;(27.8\,\mathrm{ms})$\\
PETRA III
& $3.3\;(25\,\mu\mathrm{s})$
& $2.2\times10^6\;(17.1\,\mathrm{s})$
\end{tabular}
\end{ruledtabular}
\end{table}

The comparison is the main practical result. Dipole jitter can smear a
fixed-frame image well before it changes the internal state, especially
at high energy, but this limitation is removed by event-by-event or
turn-resolved recentering. Under the stated quadrupole estimates, the
intrinsic $|\ell|=1$ sector survives of order $2\times10^5$ turns in
IOTA and $2\times10^6$ turns in PETRA III. The scaling is explicit,
\begin{align}
N_{\rm lab}^{(d)}(\ell)
&=
N_{\rm lab}^{(d)}(1)\frac{2}{|\ell|+1},
\nonumber\\
N_{\rm int}^{(q)}(\ell)
&=
N_{\rm int}^{(q)}(1)
\frac{6}{(|\ell|+1)(|\ell|+2)}.
\label{eq:main_lifetime_scaling}
\end{align}
For example, $|\ell|=10$ gives approximately
$9.5\times10^3$ intrinsic turns in IOTA and $1.0\times10^5$ in PETRA
III. This leaves a clear parameter window for preserving internal OAM:
the severe dipole requirement concerns direct observability, whereas the
quadrupole requirement concerns the state itself.

\textit{Conclusion.}
We have shown that vortex-electron transport in a storage ring is
organized by the periodic Lewis-Floquet basis of a round one-turn map that can be implemented with the Burov,
Nagaitsev, and Derbenev \cite{Burov2002} lattice.
The metaplectic lift gives the corresponding state-level transport,
while stochastic departures generate a Lindblad channel in this natural
basis. Dipole noise limits fixed-origin observability but is removed by
centroid recentering, whereas quadrupole noise produces genuine
$\Delta\ell=\pm2$ leakage. Under the stated white-noise assumption, the
$|\ell|=1$ sector survives approximately $2\times10^5$ turns on the IOTA
scale and $2\times10^6$ turns on the PETRA III scale, leaving a parameter
window in which internal OAM remains resolved after the laboratory image
has been smeared. 

We note that at high energies, synchrotron radiation adds a distinct channel that fits the same algebraic logic. To leading order, photon recoil from the reference-orbit motion acts as a conditional Weyl displacement in the energy-dependent orbital phase space. After the radiation field is traced out, the resulting random-Weyl channel smears the centroid and dispersive orbit but, upon orbit recentering, does not by itself change the intrinsic OAM distribution. Intrinsic OAM leakage arises from the non-displacement, structure-resolving part of the emission vertex. Its leading term is dipolar and permits $\Delta\ell=\pm1$ \cite{Karlovets2023,Karlovets2026}, while higher multipoles are further suppressed in the paraxial regime \cite{Epov2026}. A machine-specific decay law must therefore separate the leading Weyl-type orbital smearing from the weaker intrinsic radiative leakage.

\bibliographystyle{apsrev4-2}
\bibliography{ref}

\clearpage
\onecolumngrid

\begin{center}
{\large\bfseries Supplemental Material for\\
``State-resolved quantum transport of vortex electrons in accelerators''\par}
\vspace{1em}
S.~S.~Baturin

\vspace{0.75em}
\begin{minipage}{0.9\textwidth}
\small
The Supplemental Material contains details of the quadratic representation of the local kinetic OAM, the Lewis-Floquet invariants of the storage ring, and the construction of the metaplectic integral kernel from a symplectic matrix.
\end{minipage}
\end{center}

\setcounter{section}{0}

\section{Quadratic representation of the local kinetic OAM}
\label{app:kinetic_oam_matrix}

We derive here the explicit quadratic matrix representing the normalized
local longitudinal kinetic OAM. The centered normalized canonical
variables are ordered
as
\begin{align}
\hat Z
&=
\left(
\hat x,
\hat y,
\hat\zeta,
\hat p_x,
\hat p_y,
\hat p_\zeta
\right)^{T}.
\label{eq:app_phase_space_ordering}
\end{align}
We denote the corresponding centered coordinate vector by
$\mathbf q=(x,y,\zeta)^T$.
The normalized local kinetic OAM is
\begin{align}
\hat{\mathcal L}_s
&=
\hat x\hat\pi_y-\hat y\hat\pi_x,
&
\hat{\boldsymbol\pi}_{\perp}
&=
\hat{\mathbf p}_{\perp}
-
\chi\mathbf A_{\perp},
&
\chi
&=
\frac{q}{p_0}.
\label{eq:app_kinetic_oam_definition}
\end{align}

Expand the transverse vector potential to first order about the
reference orbit,
\begin{align}
A_x
&=
A_x^{(0)}
+
a_{xx}(s)x
+
a_{xy}(s)y
+
a_{x\zeta}(s)\zeta
+
\mathcal O(\|\mathbf q\|^2),
\\
A_y
&=
A_y^{(0)}
+
a_{yx}(s)x
+
a_{yy}(s)y
+
a_{y\zeta}(s)\zeta
+
\mathcal O(\|\mathbf q\|^2),
\label{eq:app_vector_potential_expansion}
\end{align}
where
\begin{align}
a_{ij}(s)
&=
\left.
\frac{\partial A_i}{\partial r_j}
\right|_{\mathbf R_0(s)}.
\label{eq:app_vector_potential_derivatives}
\end{align}
The constant terms $A_x^{(0)}$ and $A_y^{(0)}$ contribute only terms
linear in the centered coordinates. These terms belong to the
reference-orbit or extrinsic sector and do not enter the intrinsic
quadratic observable.

The quadratic part of the kinetic OAM is
\begin{align}
\hat{\mathcal L}_s^{(2)}
&=
\hat x\hat p_y-\hat y\hat p_x
-\chi a_{yx}\hat x^2
+\chi a_{xy}\hat y^2
\nonumber\\
&\quad
+\chi\left(a_{xx}-a_{yy}\right)\hat x\hat y
\nonumber\\
&\quad
-\chi a_{y\zeta}\hat x\hat\zeta
+\chi a_{x\zeta}\hat y\hat\zeta.
\label{eq:app_kinetic_oam_expanded}
\end{align}
It can be written as
\begin{align}
\hat{\mathcal L}_s^{(2)}
&=
\frac{1}{2}
\hat Z^{T}g_{\mathcal L}(s)\hat Z,
\label{eq:app_kinetic_oam_quadratic_form}
\end{align}
with the real symmetric matrix
\begin{align}
g_{\mathcal L}(s)
&=
\begin{pmatrix}
-2\chi a_{yx}
&
\chi(a_{xx}-a_{yy})
&
-\chi a_{y\zeta}
&
0
&
1
&
0
\\[1mm]
\chi(a_{xx}-a_{yy})
&
2\chi a_{xy}
&
\chi a_{x\zeta}
&
-1
&
0
&
0
\\[1mm]
-\chi a_{y\zeta}
&
\chi a_{x\zeta}
&
0
&
0
&
0
&
0
\\[1mm]
0
&
-1
&
0
&
0
&
0
&
0
\\[1mm]
1
&
0
&
0
&
0
&
0
&
0
\\[1mm]
0
&
0
&
0
&
0
&
0
&
0
\end{pmatrix}.
\label{eq:app_kinetic_oam_general_matrix}
\end{align}

No operator-ordering correction is required in the canonical part
because
\begin{align}
[\hat x,\hat p_y]
&=
[\hat y,\hat p_x]
=
0.
\end{align}

For a locally axisymmetric longitudinal field, the transverse vector
potential may be represented in the local symmetric gauge as
\begin{align}
\mathbf A_{\perp}
&=
\frac{B_s(s)}{2}
\left(
-y,
x
\right).
\label{eq:app_symmetric_gauge}
\end{align}
The nonzero derivatives are then
\begin{align}
a_{xy}
&=
-\frac{B_s}{2},
&
a_{yx}
&=
\frac{B_s}{2},
\label{eq:app_axisymmetric_derivatives}
\end{align}
and the kinetic OAM becomes
\begin{align}
\hat{\mathcal L}_s
&=
\hat{\mathcal L}_s^{\rm can}
-
\frac{\chi B_s(s)}{2}
\left(
\hat x^2+\hat y^2
\right),
\label{eq:app_axisymmetric_kinetic_oam}
\\
\hat{\mathcal L}_s^{\rm can}
&=
\hat x\hat p_y-\hat y\hat p_x.
\label{eq:app_canonical_oam}
\end{align}
The corresponding quadratic matrix is
\begin{align}
g_{\mathcal L}(s)
&=
\begin{pmatrix}
-\chi B_s(s)
&
0
&
0
&
0
&
1
&
0
\\
0
&
-\chi B_s(s)
&
0
&
-1
&
0
&
0
\\
0
&
0
&
0
&
0
&
0
&
0
\\
0
&
-1
&
0
&
0
&
0
&
0
\\
1
&
0
&
0
&
0
&
0
&
0
\\
0
&
0
&
0
&
0
&
0
&
0
\end{pmatrix}.
\label{eq:app_kinetic_oam_axisymmetric_matrix}
\end{align}

Restricting to the transverse phase-space ordering
\begin{align}
\hat Z_{\perp}
&=
\left(
\hat x,
\hat y,
\hat p_x,
\hat p_y
\right)^{T},
\end{align}
the same result has the compact block form
\begin{align}
g_{\mathcal L,\perp}(s)
&=
\begin{pmatrix}
-\chi B_s(s)\mathsf I_2 & \mathsf E\\
-\mathsf E & 0
\end{pmatrix},
&
\mathsf E
&=
\begin{pmatrix}
0 & 1\\
-1 & 0
\end{pmatrix}.
\label{eq:app_kinetic_oam_block_matrix}
\end{align}
Although $\mathsf E$ is antisymmetric, the full block matrix is
symmetric because the lower-left block is the transpose of the
upper-right block.

Equations~\eqref{eq:app_kinetic_oam_axisymmetric_matrix} and
\eqref{eq:app_kinetic_oam_block_matrix} also make explicit the
difference between canonical and kinetic OAM. The canonical part is
contained in the mixed position-momentum blocks, whereas the magnetic,
or diamagnetic, contribution lies in the position-position block.

The matrix representation depends on the gauge because it is expressed
in canonical phase-space variables. The complete normalized operator
$\hat{\mathcal L}_s=\bigl[\hat{\mathbf r}_{\perp}\times
(\hat{\mathbf p}_{\perp}-\chi\mathbf A_{\perp})\bigr]_s$ is nevertheless gauge
invariant: under a gauge transformation, the canonical momentum and the
matrix $g_{\mathcal L}$ transform together. The corresponding physical
operator is $\hat L_s^{\rm kin}=p_0\hat{\mathcal L}_s$.

\section{Periodic Lewis invariants and the stroboscopic basis}
\label{app:lewis_floquet_construction}

We summarize the quadratic Lewis-Riesenfeld construction for a
periodic ring Hamiltonian and its relation to the one-turn Floquet
basis. Consider
\begin{align}
\hat H_F^{(2)}(s)
&=
\frac{1}{2}
\hat Z^{T}h_F(s)\hat Z,
&
h_F(s+C)
&=
h_F(s),
\label{eq:app_periodic_quadratic_hamiltonian}
\end{align}
with
\begin{align}
\frac{d\hat Z}{ds}
&=
\mathsf A_F(s)\hat Z,
&
\mathsf A_F(s)
&=
\mathsf J_6 h_F(s).
\label{eq:app_linear_operator_evolution}
\end{align}
Let $\mathsf M(s,s_0)$ be the corresponding symplectic transport
matrix,
\begin{align}
\frac{\partial\mathsf M(s,s_0)}{\partial s}
&=
\mathsf A_F(s)\mathsf M(s,s_0),
&
\mathsf M(s_0,s_0)
&=
\mathsf I_6.
\label{eq:app_fundamental_matrix}
\end{align}

A quadratic Lewis-Riesenfeld invariant has the form
\begin{align}
\hat{\mathcal I}(s)
&=
\frac{1}{2}
\hat Z^{T}g_{\mathcal I}(s)\hat Z,
&
g_{\mathcal I}(s)
&=
g_{\mathcal I}^{T}(s).
\label{eq:app_quadratic_lewis_invariant}
\end{align}
The invariant equation
\begin{align}
\frac{\partial\hat{\mathcal I}}{\partial s}
+
\frac{\ii}{\hbar_{\rm eff}}
\left[
\hat H_F^{(2)}(s),
\hat{\mathcal I}(s)
\right]
&=
0
\label{eq:app_lewis_invariant_equation}
\end{align}
is equivalent to
\begin{align}
g_{\mathcal I}'(s)
+
\mathsf A_F^{T}(s)g_{\mathcal I}(s)
+
g_{\mathcal I}(s)\mathsf A_F(s)
&=
0.
\label{eq:app_lewis_matrix_equation}
\end{align}
For a prescribed initial quadratic form $g_{\mathcal I}(s_0)$, the solution is
\begin{align}
g_{\mathcal I}(s)
&=
\mathsf M^{-T}(s,s_0)
g_{\mathcal I}(s_0)
\mathsf M^{-1}(s,s_0).
\label{eq:app_lewis_matrix_solution}
\end{align}
Equivalently, at the operator level,
\begin{align}
\hat{\mathcal I}(s)
&=
\hat U(s,s_0)
\hat{\mathcal I}(s_0)
\hat U^{\dagger}(s,s_0).
\label{eq:app_lewis_operator_solution}
\end{align}
The operator is therefore explicitly $s$ dependent when expressed in
the local canonical variables, whereas its spectrum and its
expectation value in an exactly transported state remain constant.

Let
\begin{align}
\mathsf M_C
&=
\mathsf M(s_0+C,s_0)
\label{eq:app_one_turn_matrix}
\end{align}
denote the one-turn map. Periodicity of the Hamiltonian matrix gives
\begin{align}
\mathsf M(s+C,s_0)
&=
\mathsf M(s,s_0)\mathsf M_C.
\label{eq:app_periodic_fundamental_matrix}
\end{align}
Substitution into Eq.~\eqref{eq:app_lewis_matrix_solution} shows that
the invariant is periodic,
\begin{align}
g_{\mathcal I}(s+C)
&=
g_{\mathcal I}(s),
\label{eq:app_periodic_lewis_matrix}
\end{align}
if and only if its initial quadratic form satisfies
\begin{align}
\mathsf M_C^{T}
g_{\mathcal I}(s_0)
\mathsf M_C
&=
g_{\mathcal I}(s_0).
\label{eq:app_one_turn_quadratic_invariant}
\end{align}
At the operator level, this condition becomes
\begin{align}
\hat U_C^{\dagger}
\hat{\mathcal I}(s_0)
\hat U_C
&=
\hat{\mathcal I}(s_0),
&
\hat U_C
&=
\hat U(s_0+C,s_0).
\label{eq:app_one_turn_operator_invariant}
\end{align}
Thus a periodic Lewis-Riesenfeld invariant and a stroboscopic
one-turn invariant are the continuous and fixed-azimuth
representations of the same structure.

We now restrict the construction to the invariant transverse
subspace of a stable round map. Introduce normalized canonical
variables
\begin{align}
\hat Z_n
&=
\left(
\hat X,
\hat Y,
\hat P_X,
\hat P_Y
\right)^{T}.
\label{eq:app_normalized_transverse_variables}
\end{align}
The two normalized quadratic operators relevant to the vortex basis
are the total oscillator action
\begin{align}
\hat{\mathcal I}_{0,n}
&=
\frac{1}{2}
\left(
\hat X^2+\hat Y^2+\hat P_X^2+\hat P_Y^2
\right)
=
\frac{1}{2}
\hat Z_n^{T}g_{0,n}\hat Z_n,
\label{eq:app_normalized_total_action}
\end{align}
and the normalized OAM generator
\begin{align}
\hat L_n
&=
\hat X\hat P_Y-\hat Y\hat P_X
=
\frac{1}{2}
\hat Z_n^{T}g_{L,n}\hat Z_n,
\label{eq:app_normalized_oam_generator}
\end{align}
where
\begin{align}
g_{0,n}
&=
\mathsf I_4,
&
g_{L,n}
&=
\begin{pmatrix}
0 & \mathsf E\\
-\mathsf E & 0
\end{pmatrix},
&
\mathsf E
&=
\begin{pmatrix}
0 & 1\\
-1 & 0
\end{pmatrix}.
\label{eq:app_normalized_invariant_matrices}
\end{align}
These operators commute,
\begin{align}
\left[
\hat{\mathcal I}_{0,n},
\hat L_n
\right]
&=
0.
\label{eq:app_normalized_invariant_commutation}
\end{align}

Let $\mathsf N(s_0)$ denote the symplectic transformation from the
local transverse accelerator variables to the Floquet-normalized
variables at the reference azimuth,
\begin{align}
\hat Z_n
&=
\mathsf N^{-1}(s_0)\hat Z_{\perp}.
\label{eq:app_floquet_normalization}
\end{align}
The normalized one-turn map is then
\begin{align}
\mathsf M_{n,C}
&=
\mathsf N^{-1}(s_0)
\mathsf M_{\perp,C}
\mathsf N(s_0),
\label{eq:app_normalized_one_turn_map}
\end{align}
where $\mathsf M_{\perp,C}$ is the restriction of the one-turn map to
the invariant transverse subspace. The stable Floquet normalization
and the round-map condition imply, respectively,
\begin{align}
\mathsf M_{n,C}^{T}
g_{0,n}
\mathsf M_{n,C}
&=
g_{0,n},
\nonumber\\
\mathsf M_{n,C}^{T}
g_{L,n}
\mathsf M_{n,C}
&=
g_{L,n}.
\label{eq:app_normalized_one_turn_invariants}
\end{align}

The corresponding quadratic forms in the local variables are
\begin{align}
g_{\mathcal I_a}(s_0)
&=
\mathsf N^{-T}(s_0)
g_{a,n}
\mathsf N^{-1}(s_0),
&
a
&=
0,L.
\label{eq:app_local_invariant_initial_matrices}
\end{align}
They satisfy
\begin{align}
\mathsf M_{\perp,C}^{T}
g_{\mathcal I_a}(s_0)
\mathsf M_{\perp,C}
&=
g_{\mathcal I_a}(s_0),
&
a
&=
0,L,
\label{eq:app_local_one_turn_invariant_conditions}
\end{align}
and therefore generate the two periodic Lewis invariants
\begin{align}
\hat{\mathcal I}_a(s)
&=
\frac{1}{2}
\hat Z_{\perp}^{T}
g_{\mathcal I_a}(s)
\hat Z_{\perp},
\nonumber\\
g_{\mathcal I_a}(s)
&=
\mathsf M_{\perp}^{-T}(s,s_0)
g_{\mathcal I_a}(s_0)
\mathsf M_{\perp}^{-1}(s,s_0),
\qquad
a=0,L.
\label{eq:app_periodic_transverse_lewis_invariants}
\end{align}
Since both operators are transported by the same unitary evolution,
their commutation relation is preserved:
\begin{align}
\left[
\hat{\mathcal I}_0(s),
\hat{\mathcal I}_L(s)
\right]
&=
0.
\label{eq:app_total_action_oam_commutation}
\end{align}

The operator $\hat{\mathcal I}_L(s)$ is the continuously transported
local-coordinate representative of the normalized OAM generator. It
should not, in general, be identified with the instantaneous local
kinetic OAM $\hat{\mathcal L}_s(s)$. The second invariant
$\hat{\mathcal I}_0(s)$ resolves the remaining transverse degeneracy
and completes the definition of the vortex basis.

We define their common eigenstates by
\begin{align}
\hat{\mathcal I}_0(s)
\lvert\phi_{n\ell}(s)\rangle
&=
\hbar_{\rm eff}
\left(
2n+|\ell|+1
\right)
\lvert\phi_{n\ell}(s)\rangle,
\nonumber\\
\hat{\mathcal I}_L(s)
\lvert\phi_{n\ell}(s)\rangle
&=
\hbar_{\rm eff}\ell
\lvert\phi_{n\ell}(s)\rangle,
\label{eq:app_total_action_oam_eigenstates}
\end{align}
where
\begin{align}
n
&\in
\mathbb N_0,
&
\ell
&\in
\mathbb Z.
\end{align}
The label $n$ resolves the radial degeneracy at fixed $\ell$, but no
separate radial-action operator is required. The commuting pair
$\hat{\mathcal I}_0(s)$ and $\hat{\mathcal I}_L(s)$ therefore provides
the basis labels used in the main text.

At the reference azimuth, both invariants commute with the one-turn
propagator,
\begin{align}
\left[
\hat U_C,
\hat{\mathcal I}_a(s_0)
\right]
&=
0,
&
a
&=
0,L.
\label{eq:app_invariants_commute_with_map}
\end{align}
Their common eigenvectors can consequently be chosen as one-turn
Floquet eigenvectors,
\begin{align}
\hat U_C
\lvert n,\ell;s_0\rangle_L
&=
\ee^{-\ii\vartheta_{n\ell}}
\lvert n,\ell;s_0\rangle_L.
\label{eq:app_stroboscopic_lewis_basis}
\end{align}
The joint $(n,\ell)$ eigenspace is one-dimensional within the
transverse sector. If additional degrees of freedom produce a
degeneracy, Eq.~\eqref{eq:app_stroboscopic_lewis_basis} is understood
after diagonalizing $\hat U_C$ within the corresponding joint
invariant subspace. The resulting vectors define the stroboscopic
Lewis-Floquet basis.

The continuous local basis is formed by the instantaneous
eigenvectors of the periodic invariants. Their phases may be chosen
such that
\begin{align}
\lvert\phi_{n\ell}(s+C)\rangle
&=
\lvert\phi_{n\ell}(s)\rangle.
\label{eq:app_periodic_lewis_basis}
\end{align}
The corresponding exact transported solution is
\begin{align}
\lvert\psi_{n\ell}(s)\rangle
&=
\ee^{\ii\alpha_{n\ell}(s)}
\lvert\phi_{n\ell}(s)\rangle,
\label{eq:app_lewis_solution}
\end{align}
where the Lewis-Riesenfeld phase obeys
\begin{align}
\frac{d\alpha_{n\ell}}{ds}
&=
\left\langle
\phi_{n\ell}(s)
\left|
\ii\frac{\partial}{\partial s}
-
\frac{\hat H_F^{(2)}(s)}{\hbar_{\rm eff}}
\right|
\phi_{n\ell}(s)
\right\rangle.
\label{eq:app_lewis_phase}
\end{align}
Choosing $\alpha_{n\ell}(s_0)=0$, after one circumference one obtains
\begin{align}
\hat U_C
\lvert\phi_{n\ell}(s_0)\rangle
&=
\ee^{\ii\alpha_{n\ell}(s_0+C)}
\lvert\phi_{n\ell}(s_0)\rangle.
\label{eq:app_lewis_floquet_phase_relation}
\end{align}
The accumulated Lewis-Riesenfeld phase is therefore the Floquet
eigenphase of the one-turn map. With the convention used in
Eq.~\eqref{eq:app_stroboscopic_lewis_basis},
\begin{align}
\vartheta_{n\ell}
&=
-\alpha_{n\ell}(s_0+C)
\pmod{2\pi}.
\label{eq:app_floquet_phase_sign_relation}
\end{align}

The same structure follows from the Floquet decomposition
\begin{align}
\hat U(s,s_0)
&=
\hat P(s,s_0)
\exp\left[
-\frac{\ii}{\hbar_{\rm eff}}
(s-s_0)\hat K_F
\right],
\nonumber\\
\hat P(s+C,s_0)
&=
\hat P(s,s_0),
\label{eq:app_floquet_decomposition}
\end{align}
where $\hat P$ describes the periodic intra-turn micromotion and
$\hat K_F$ is the quadratic Floquet generator. If a quadratic
operator $\hat{\mathcal J}_F$ commutes with $\hat K_F$, then
\begin{align}
\hat{\mathcal J}(s)
&=
\hat P(s,s_0)
\hat{\mathcal J}_F
\hat P^{\dagger}(s,s_0)
\label{eq:app_floquet_lewis_invariant}
\end{align}
is a periodic Lewis-Riesenfeld invariant. The periodic basis profiles
are generated by the micromotion operator $\hat P$, whereas the
eigenvalues of $\hat K_F$ determine the one-turn phases.

For a matched Lewis-Floquet mode, the probability profile and all
quadratic moments are periodic around the ring. An unmatched injected
state is instead a squeezed or sheared superposition of the matched
Lewis-Floquet modes. The existence of the invariants and of the
stroboscopic basis is determined by the ideal one-turn map and does
not require imposing covariance matching on an arbitrary transported
state.

\section{Metaplectic propagator as a Gaussian integral operator}
\label{app:metaplectic_kernel}

We record the coordinate-space representation of the metaplectic
propagator associated with a linear symplectic transfer map. This
provides the state-level counterpart of the covariance transport law.

Let the canonical variables relative to the reference orbit be
\begin{align}
\hat Z
&=
\begin{pmatrix}
\hat{\mathbf q}\\
\hat{\mathbf p}
\end{pmatrix},
&
\left[
\hat q_i,\hat p_j
\right]
&=
\ii\hbar_{\rm eff}\delta_{ij},
\label{eq:app_canonical_coordinate_vector}
\end{align}
where $\mathbf q\in\mathbb R^d$. In the present application, one may
take $d=2$ for the transverse dynamics or $d=3$ for the full
$(x,y,\zeta)$ system. A quadratic Hamiltonian generates the symplectic
transfer matrix
\begin{align}
\mathsf M(s,s_0)
&=
\begin{pmatrix}
\mathsf A & \mathsf B\\
\mathsf C & \mathsf D
\end{pmatrix},
&
\mathsf M^{T}\mathsf J_{2d}\mathsf M
&=
\mathsf J_{2d}.
\label{eq:app_symplectic_blocks}
\end{align}
The corresponding classical transfer relation is
\begin{align}
\begin{pmatrix}
\mathbf q\\
\mathbf p
\end{pmatrix}
&=
\mathsf M(s,s_0)
\begin{pmatrix}
\mathbf q_0\\
\mathbf p_0
\end{pmatrix}.
\label{eq:app_classical_transfer_relation}
\end{align}

The quantum evolution is represented by the metaplectic lift
$\hat U(s,s_0)$ of $\mathsf M(s,s_0)$, satisfying
\begin{align}
\hat U^{\dagger}(s,s_0)
\hat Z
\hat U(s,s_0)
&=
\mathsf M(s,s_0)\hat Z.
\label{eq:app_metaplectic_defining_relation}
\end{align}
The symplectic matrix alone determines two metaplectic lifts differing
by an overall sign. For a continuous Hamiltonian evolution, the path
from $s_0$ to $s$ fixes the lift and its Maslov index. The covariance
transport law
\begin{align}
\Sigma(s)
&=
\mathsf M(s,s_0)
\Sigma(s_0)
\mathsf M^{T}(s,s_0)
\label{eq:app_covariance_metaplectic_transport}
\end{align}
is the second-moment projection of this unitary evolution.

Assume first that the block $\mathsf B$ is invertible. The symplectic
map then admits a type-I generating function
$S_{\mathsf M}(\mathbf q,\mathbf q_0)$. From
\begin{align}
\mathbf q
&=
\mathsf A\mathbf q_0+\mathsf B\mathbf p_0,
\end{align}
one obtains
\begin{align}
\mathbf p_0
&=
\mathsf B^{-1}
\left(
\mathbf q-\mathsf A\mathbf q_0
\right).
\end{align}
Using the symplectic identities, the final momentum becomes
\begin{align}
\mathbf p
&=
\mathsf D\mathsf B^{-1}\mathbf q
-
\mathsf B^{-T}\mathbf q_0.
\end{align}
The corresponding generating function is
\begin{align}
S_{\mathsf M}(\mathbf q,\mathbf q_0)
&=
\frac{1}{2}
\mathbf q^{T}
\mathsf D\mathsf B^{-1}
\mathbf q
-
\mathbf q^{T}
\mathsf B^{-T}
\mathbf q_0
\nonumber\\
&\quad
+
\frac{1}{2}
\mathbf q_0^{T}
\mathsf B^{-1}\mathsf A
\mathbf q_0.
\label{eq:app_metaplectic_generating_function}
\end{align}
Indeed,
\begin{align}
\mathbf p
&=
\frac{\partial S_{\mathsf M}}{\partial\mathbf q},
&
\mathbf p_0
&=
-
\frac{\partial S_{\mathsf M}}{\partial\mathbf q_0}.
\end{align}
The matrices $\mathsf D\mathsf B^{-1}$ and
$\mathsf B^{-1}\mathsf A$ are symmetric as a consequence of the
symplectic conditions.

The coordinate-space propagator is the Gaussian kernel
\begin{align}
K_{\mathsf M}
(\mathbf q,\mathbf q_0;s,s_0)
&=
\frac{
\ee^{\ii\chi_{\mathsf M}}
}{
(2\pi\hbar_{\rm eff})^{d/2}
|\det\mathsf B|^{1/2}
}
\exp
\left[
\frac{\ii}{\hbar_{\rm eff}}
S_{\mathsf M}(\mathbf q,\mathbf q_0)
\right].
\label{eq:app_metaplectic_kernel_general}
\end{align}
Here $\chi_{\mathsf M}$ contains the Fresnel and Maslov phases associated
with the chosen metaplectic lift. It is fixed within a given
metaplectic chart and changes when the transport crosses a caustic,
where $\det\mathsf B=0$. This phase cancels from density-matrix
evolution along a single path but must be retained when amplitudes associated with different
paths or metaplectic charts interfere coherently.

The wavefunction evolves according to
\begin{align}
\psi(\mathbf q,s)
&=
\int d^d q_0\,
K_{\mathsf M}
(\mathbf q,\mathbf q_0;s,s_0)
\psi(\mathbf q_0,s_0).
\label{eq:app_wavefunction_metaplectic_transport}
\end{align}
Writing
\begin{align}
\rho(\mathbf q,\mathbf q';s)
&=
\left\langle
\mathbf q
\left|
\rho(s)
\right|
\mathbf q'
\right\rangle,
\end{align}
the corresponding density-matrix transport is
\begin{align}
\rho(\mathbf q,\mathbf q';s)
&=
\int d^d q_0\,d^d q_0'\,
K_{\mathsf M}
(\mathbf q,\mathbf q_0;s,s_0)
\nonumber\\
&\quad\times
\rho(\mathbf q_0,\mathbf q_0';s_0)
K_{\mathsf M}^{*}
(\mathbf q',\mathbf q_0';s,s_0).
\label{eq:app_density_matrix_kernel_transport}
\end{align}

For one transverse degree of freedom, the blocks reduce to scalars and
$\mathsf M$ becomes the usual ABCD matrix,
\begin{align}
\mathsf M
&=
\begin{pmatrix}
A&B\\
C&D
\end{pmatrix},
&
AD-BC
&=
1.
\end{align}
For $B\neq0$, Eq.~\eqref{eq:app_metaplectic_kernel_general} reduces to
\begin{align}
K_{ABCD}(x,x_0)
&=
\frac{
\ee^{\ii\chi_{\mathsf M}}
}{
\sqrt{2\pi\hbar_{\rm eff}|B|}
}
\nonumber\\
&\quad\times
\exp
\left[
\frac{\ii}{2\hbar_{\rm eff}B}
\left(
D x^2-2xx_0+A x_0^2
\right)
\right].
\label{eq:app_abcd_kernel}
\end{align}
This is the Gaussian Green function familiar from paraxial wave optics,
with $\hbar_{\rm eff}$ playing the role of the effective paraxial
commutator. The phase $\chi_{\mathsf M}$ includes the branch associated
with the sign of $B$.

When $\mathsf B$ is singular,
Eq.~\eqref{eq:app_metaplectic_kernel_general} must be replaced by another
metaplectic chart. Equivalently, the symplectic matrix may be factored
into elementary shear, scaling, and Fourier-transform maps. In the
special case $\mathsf B=0$, the symplectic matrix has the form
\begin{align}
\mathsf M
&=
\begin{pmatrix}
\mathsf A&0\\
\mathsf C&\mathsf A^{-T}
\end{pmatrix}.
\end{align}
The metaplectic operator then acts as
\begin{align}
\left(
\hat U\psi
\right)(\mathbf q)
&=
\frac{
\ee^{\ii\chi_{\mathsf M}}
}{
|\det\mathsf A|^{1/2}
}
\exp
\left[
\frac{\ii}{2\hbar_{\rm eff}}
\mathbf q^{T}
\mathsf C\mathsf A^{-1}
\mathbf q
\right]
\nonumber\\
&\quad\times
\psi(\mathsf A^{-1}\mathbf q),
\label{eq:app_B_zero_metaplectic}
\end{align}
where $\mathsf C\mathsf A^{-1}$ is symmetric. Equivalently, its kernel is
\begin{align}
K_{\mathsf M}(\mathbf q,\mathbf q_0)
&=
\frac{
\ee^{\ii\chi_{\mathsf M}}
}{
|\det\mathsf A|^{1/2}
}
\exp
\left[
\frac{\ii}{2\hbar_{\rm eff}}
\mathbf q^{T}
\mathsf C\mathsf A^{-1}
\mathbf q
\right]
\nonumber\\
&\quad\times
\delta^{(d)}
\left(
\mathbf q_0-\mathsf A^{-1}\mathbf q
\right).
\label{eq:app_B_zero_kernel}
\end{align}

If the quadratic Hamiltonian also contains linear terms, the classical
transport becomes affine,
\begin{align}
Z(s)
&=
\mathsf M(s,s_0)Z(s_0)
+
Z_c(s,s_0),
\label{eq:app_affine_classical_transport}
\end{align}
where $Z_c(s,s_0)$ is the inhomogeneous classical trajectory with
$Z_c(s_0,s_0)=0$. Define the Weyl displacement operator by
\begin{align}
\hat W(Z_c)
&=
\exp
\left[
-\frac{\ii}{\hbar_{\rm eff}}
Z_c^{T}\mathsf J_{2d}\hat Z
\right],
&
\hat W^{\dagger}(Z_c)
\hat Z
\hat W(Z_c)
&=
\hat Z+Z_c.
\label{eq:app_weyl_displacement}
\end{align}
The affine quantum propagator can then be written as
\begin{align}
\hat U_{\rm aff}(s,s_0)
&=
\ee^{\ii\phi(s,s_0)}
\hat W[Z_c(s,s_0)]
\hat U(s,s_0),
\label{eq:app_affine_metaplectic}
\end{align}
where $\phi(s,s_0)$ is a scalar phase. It follows that
\begin{align}
\hat U_{\rm aff}^{\dagger}
\hat Z
\hat U_{\rm aff}
&=
\mathsf M(s,s_0)\hat Z
+
Z_c(s,s_0).
\end{align}
After recentering on the transported centroid, the internal density
matrix is therefore governed by the homogeneous metaplectic propagator.
A deterministic dipole perturbation produces a removable Weyl
displacement, while stochastic dipole jitter generates an ensemble of
such centroid displacements.

\end{document}